\documentstyle[aps,prl,floats]{revtex}
\begin{document}
\draft
\tighten
\twocolumn[\hsize\textwidth\columnwidth\hsize\csname@twocolumnfalse%
\endcsname
\title{High Energy Neutrinos from Cosmological Gamma-Ray Burst Fireballs}
\author{Eli Waxman and John Bahcall}
\address{Institute for Advanced Study, Princeton, NJ 08540; E-mail:
waxman@sns.ias.edu}
\date{\today}
\maketitle
\begin{abstract}
Observations suggest that $\gamma$-ray bursts (GRBs) 
are produced by the dissipation
of the kinetic energy of a relativistic fireball. We show that a large
fraction, $\ge10\%$, of the fireball energy is expected to be converted
by photo-meson production to a burst of 
$\sim10^{14}{\rm eV}$ neutrinos. A ${\rm km}^2$ neutrino detector would
observe at least several tens of events per year correlated with GRBs, and
test for neutrino properties (e.g. flavor oscillations,
for which upward moving $\tau$'s would be a unique signature, and coupling
to gravity) with an accuracy many orders 
of magnitude better than is currently possible.
\end{abstract}

]

\narrowtext

Recent observations of $\gamma$-ray bursts (GRBs)
suggest that they originate from 
cosmological sources \cite{cos} (see, however, \cite{galactic}). 
General phenomenological considerations indicate that the 
bursts are produced by the dissipation of the kinetic energy of a 
relativistic expanding fireball (see \cite{scenario} for reviews). 
The physical conditions in the dissipation region imply \cite{CRprod} that 
protons may be Fermi accelerated in 
this region to energies $>10^{20}{\rm eV}$. Furthermore, the spectrum and flux
of ultra-high energy cosmic rays (above
$10^{19}{\rm eV}$) are consistent with those expected from Fermi acceleration
of protons in cosmological GRBs \cite{CRflux}. 
We show in this Letter that a natural consequence of the dissipative fireball
model of GRBs is the conversion of a significant fraction
of the fireball energy to an accompanying burst of $\sim10^{14}{\rm eV}$ 
neutrinos, created by photo-meson production of pions in
interactions between the fireball $\gamma$-rays and accelerated protons. 
The neutrino burst is produced by interaction
with protons with energies much lower than $\sim10^{20}{\rm eV}$,
the maximum acceleration energy. As shown below, $10^{15}{\rm eV}$ protons
interact with the
$\sim1{\rm MeV}$ photons carrying the bulk of $\gamma$-ray energy to
produce $\sim10^{14}{\rm eV}$ neutrinos.

The rapid variability time, $\sim1\text{ms}$, observed in some GRBs implies
that the sources are compact, with a linear scale $r_0\sim10^7\text{cm}$. 
The high luminosity required for cosmological bursts, 
$\sim10^{51}{\rm erg \ s}^{-1}$,
then results in an optically thick (to pair creation) plasma,
which expands and accelerates to 
relativistic velocities \cite{fire}. The hardness of the observed
photon spectra, which extends to $\ge100\text{MeV}$, implies 
that the $\gamma$-ray emitting region must be moving  
with a Lorentz factor $\Gamma$ of order $100$ \cite{lorentz}, and 
constitutes independent evidence for ultra-relativistic outflow.
The high energy density in the source would result in complete thermalization 
and a black-body spectrum \cite{fire}, in contrast with observations.
To overcome this problem, Rees and M\'esz\'aros suggested \cite{coll} that 
$\gamma$-ray emission results from the dissipation at large radius 
of the kinetic energy of the relativistic ejecta. Such dissipation 
is expected to occur due to a collision
with the inter-stellar medium \cite{coll}, or due to 
internal collisions within the ejecta \cite{Xu,RnM}.

Paczy\'nski and Xu suggested \cite{Xu}, that $\gamma$-rays are emitted by 
the decay of neutral pions, which are produced in $pp$ collisions
once the kinetic energy is dissipated through internal collisions.
In this case, an accompanying burst
of $\sim30{\rm GeV}$ neutrinos is expected due to the decay of charged pions.
It is not clear, however, whether this model can account for
the observed spectra. A more conservative 
mechanism for $\gamma$-ray production
is the emission of synchrotron radiation (possibly followed by
inverse-Compton scattering) by relativistic electrons accelerated in the
dissipation shocks \cite{coll,RnM}. In this case, the proton density in 
the wind is too low to allow significant conversion of energy to neutrinos
through $pp$ collisions.

In the region where electrons are accelerated, protons are also expected to be
shock accelerated. This is similar to what is thought to occur in supernovae 
remnant shocks, where synchrotron radiation of accelerated electrons is the
likely source of non-thermal X-rays (recent ASCA observations give evidence
for acceleration of electrons in the remnant of SN1006 to $10^{14}{\rm eV}$ 
\cite{SN1006}), and where shock acceleration of protons is believed to
produce cosmic rays with energy extending to $\sim10^{15}{\rm eV}$ (see, e.g.
\cite{Bland} for review). The spectrum of ultra-high energy 
cosmic rays (above $10^{19}{\rm eV}$) is consistent \cite{CRflux} 
with that expected from Fermi acceleration
of protons in cosmological GRBs, and the flux is consistent with this scenario
\cite{CRprod} provided the efficiency with which kinetic energy is converted to
accelerated protons is comparable to
the efficiency with which energy is converted to accelerated 
electrons (and hence to $\gamma$-rays). 
We derive below the expected spectrum and flux of high energy neutrinos,
produced by photo-meson interactions between the wind $\gamma$-rays and 
shock-accelerated protons, 
and discuss the implications for high energy neutrino astronomy.

\paragraph*{Neutrino production in dissipative wind models of GRBs.}

We consider a compact source producing a wind,
characterized by an average luminosity $L\sim10^{51}{\rm erg\ s}^{-1}$
and mass loss rate $\dot M=L/\eta c^2$. At small radius, 
the wind bulk Lorentz factor, $\Gamma$, 
grows linearly with radius, until most of the wind energy is converted
to kinetic energy and $\Gamma$ saturates at $\Gamma\sim\eta\sim100$.
Variability of the source on time scale $\Delta t$, resulting
in fluctuations in the wind bulk Lorentz factor $\Gamma$ on similar
time scale, would lead to internal shocks in the ejecta at a radius
$r\sim r_d\approx\Gamma^2c\Delta t$. 
We assume that internal shocks reconvert a substantial 
part of the kinetic energy to internal energy, which is then radiated as 
$\gamma$-rays by synchrotron and inverse-Compton radiation of
shock-accelerated electrons.

The photon distribution in the wind rest frame is isotropic. Denoting
by $n_\gamma(\epsilon_\gamma)d\epsilon_\gamma$ the
number density of photons in the energy range $\epsilon_\gamma$
to $\epsilon_\gamma+d\epsilon_\gamma$ in the wind rest frame, 
the fractional energy loss rate of
a proton with energy $\epsilon_p$ in the wind rest frame due to
pion production is
\begin{eqnarray}
t_\pi^{-1}(\epsilon_p)\equiv&&
-{1\over\epsilon_p}{d\epsilon_p\over dt}\nonumber\\=&&
{1\over2\Gamma_p^2}c\int_{\epsilon_0}^\infty 
d\epsilon\,\sigma_\pi(\epsilon)
\xi(\epsilon)\epsilon\,\int_{\epsilon/2\Gamma_p}^\infty dx\,
x^{-2}n(x)\,,
\label{pirate}
\end{eqnarray}
where $\Gamma_p=\epsilon_p/m_pc^2$, $\sigma_\pi(\epsilon)$ is the cross
section for pion production for a photon with energy $\epsilon$ in the
proton rest frame, $\xi(\epsilon)$ is the average fraction of energy
lost to the pion, and $\epsilon_0=0.15{\rm GeV}$ is the threshold
energy. The GRB photon spectrum is well fitted in the
BATSE range ($30$~KeV--$3$~MeV) by a combination
of two power-laws, $n(\epsilon_\gamma)\propto\epsilon_\gamma^{-\beta}$ 
with different values of $\beta$ at low and high energy \cite{Band}. The
break energy (where $\beta$ changes) in the observer frame is typically 
$\epsilon^{\rm ob.}_{\gamma b}\sim1{\rm MeV}$, 
with $\beta\simeq1$ at energies below the break and $\beta\simeq2$ 
above the break. Hereafter we denote quantities measured in the observer
frame with the super-script ``ob.'' (e.g., $\epsilon^{\rm ob.}_{\gamma b}
=\Gamma\epsilon_{\gamma b}$).
The second integral in (\ref{pirate}) may be approximated by
\begin{equation}
\int_\epsilon^\infty{\rm d}x\,x^{-2}n(x)\simeq
{1\over1+\beta}{U_\gamma\over2\epsilon_{\gamma b}^3}
\left({\epsilon\over\epsilon_{\gamma b}}\right)^{-(1+\beta)}\,,
\end{equation}
where $U_\gamma$ is the photon energy density (in the range
corresponding to the observed BATSE range) in the wind rest-frame,
$\beta=1$ for $\epsilon<\epsilon_{\gamma b}$ and 
$\beta=2$ for $\epsilon>\epsilon_{\gamma b}$.
The main contribution to the first integral in (\ref{pirate}) is from 
photon energies $\epsilon\sim\epsilon_{\rm peak}=
0.3{\rm GeV}$, where the cross section
peaks due to the $\Delta$ resonance. Approximating the integral by the
contribution from the resonance we obtain
\begin{equation}
t_\pi^{-1}(\epsilon_p)\simeq
{U_\gamma\over2\epsilon_{\gamma b}}c\sigma_{\rm peak}\xi_{\rm peak}
{\Delta\epsilon\over\epsilon_{\rm peak}}
\min(1,2\Gamma_p\epsilon_{\gamma b}/\epsilon_{\rm peak})\,.
\label{taupi}
\end{equation}
Here, $\sigma_{\rm peak}\simeq5\times10^{-28}{\rm cm}^2$ and
$\xi_{\rm peak}\simeq0.2$ are the values of $\sigma$ and $\xi$
at $\epsilon=\epsilon_{\rm peak}$, and $\Delta\epsilon\simeq
0.2{\rm GeV}$ is the peak width.

The energy loss of protons due to pion production is small during the
acceleration process \cite{CRprod}. Once accelerated, the time available
for proton energy loss by pion production is comparable to
the wind expansion time as measured in the wind rest frame, $t_d\sim r_d/
\Gamma c$. Thus, the fraction of energy lost by protons to pions is 
$f_\pi\simeq r_d/\Gamma ct_\pi$. 
The energy density in the BATSE range, $U_\gamma$, is related to the
luminosity $L_\gamma$ by $L_\gamma=4\pi r_d^2\Gamma^2cU_\gamma$. 
Using this relation in (\ref{taupi}), $f_\pi$ is given by
\begin{eqnarray}
f_\pi(\epsilon_p^{\rm ob.})=&&0.20{L_{\gamma,51}\over
\epsilon_{\gamma b,{\rm MeV}}^{\rm ob.}\Gamma_{300}^4 \Delta t_{\rm ms}}
\nonumber\\&&\times
\cases{1,&if $\epsilon_{p}^{\rm ob.}>\epsilon_{pb}^{\rm ob.}$;\cr
\epsilon_{p}^{\rm ob.}/\epsilon_{pb}^{\rm ob.},&otherwise.\cr}
\label{fpi}
\end{eqnarray}
Here, $L_\gamma=10^{51}L_{\gamma,51}{\rm erg\ s}^{-1}$,
$\Gamma=300\Gamma_{300}$, $\Delta t=10^{-3}\Delta t_{\rm ms}{\rm s}$,
and the proton break energy is
\begin{equation}
\epsilon_{pb}^{\rm ob.}=1.3\times10^{16}\Gamma_{300}^2
(\epsilon_{\gamma b,{\rm MeV}}^{\rm ob.})^{-1}\, 
{\rm eV}\,.
\label{epb}
\end{equation}
Thus, for parameters typical of a GRB producing wind, a significant fraction
of the energy of protons accelerated to energies larger than
the break energy, $\sim10^{16}{\rm eV}$, would be lost to pion production.
Note, that since the flow is ultra-relativistic, the results given above 
are independent
of whether the wind is spherically symmetric or jet-like, provided the jet 
opening angle is $>1/\Gamma$ (for a jet-like wind, $L$ is the luminosity that
would have been produced by the wind if it were spherically symmetric).

Since the constraint on the bulk
Lorentz factor, $\Gamma\ge100$, is derived from the requirement that
the wind be optically thin to pair production for photons with observed
energy $\sim100{\rm MeV}$, it is useful to express $f_\pi$ as a function
of the pair production optical depth $\tau_{\gamma\gamma}$.
A test photon with energy $\epsilon_{\rm t}$ may
produce pairs in interactions with photons with energy exceeding 
a threshold $\epsilon_{\rm th}$, determined by 
$\epsilon_{\rm t}\epsilon_{\rm th}=2(m_ec^2)^2/(1-\cos\theta)$,
where $\theta$ is the angle between the photons propagation directions.
Pair production interactions involve photons with energy much higher than
the break energy, $\epsilon_{\gamma b}$.
A test photon with observed energy $\epsilon_{\rm t}^{\rm ob.}=100
{\rm\ MeV}$, for example,
has an energy $\sim1{\rm MeV}$ in the wind rest frame, and
therefore interacts mainly with photons of similar energy to produce pairs.
Assuming that the spectrum of photons produced in the dissipation region
extends as $n(\epsilon)\propto\epsilon^{-2}$ from the BATSE range 
to (observed) energies well above $100{\rm MeV}$, the mean free path for pair
production (in the wind rest frame) for a photon of energy 
$\epsilon_{\rm t}$ is
\begin{eqnarray}
l_{\gamma\gamma}^{-1}(\epsilon_{\rm t})
&&={1\over2}{3\over16}\sigma_T\int d\cos\theta(1-\cos\theta)
\int_{\epsilon_{\rm th}(\epsilon_{\rm t},\theta)}^\infty d\epsilon
{U_\gamma\over2\epsilon^2}\nonumber\\&&
={1\over16}\sigma_T{U_\gamma\epsilon_{\rm t}\over(m_ec^2)^2} \,.
\label{tau}
\end{eqnarray}
Here we have used a constant cross section, $3\sigma_T/16$ where
$\sigma_T$ is the Thomson cross section, above the threshold $\epsilon_
{\rm th}$ [The cross section drops as $\log(\epsilon)/\epsilon$ for
$\epsilon\gg\epsilon_{\rm th}$; however, since the number density of
photons drops rapidly with energy, (\ref{tau}) is a good approximation].

The optical depth is given by $\tau_{\gamma\gamma}=
r_d/\Gamma l_{\gamma\gamma}$. Using (\ref{tau}), we obtain
\begin{equation}
f_\pi(\epsilon_p^{\rm ob.})=0.15\tau_{\gamma\gamma}
(\epsilon_{\rm t}^{\rm ob.}=100{\rm MeV}){\min(\epsilon_p^{\rm ob.},
\epsilon_{pb}^{\rm ob.})\over10^{16}{\rm eV}}\,.
\label{fpi1}
\end{equation}
Only a small fraction of bursts show a power law spectrum
extending to $>100{\rm MeV}$. Most bursts may therefore have
$\tau_{\gamma\gamma}(\epsilon_{\rm t}^{\rm ob.}=100{\rm MeV})$ 
larger than unity, leading to higher efficiency of pion 
production. Furthermore, variability of the source
over different time scales would lead to dissipation shocks over a range
of radii \cite{RnM}, where the optical depth may become
small only at the largest dissipation radii. 
While photons can escape 
only from radii where the optical depth is small, neutrinos can escape from
almost any depth. Thus, most of the burst energy may actually come out in
neutrinos. For $\Gamma=100$ and 
$L_\gamma=10^{51}{\rm erg\ s}^{-1}$, for example,
$\tau_{\gamma\gamma}
(\epsilon_{\rm t}^{\rm ob.}=100{\rm MeV})\sim1$ for $r_d\sim10^{14}{\rm cm}$,
where internal shocks result from variability over $0.1{\rm s}$ time 
scale. Variability on shorter time scales results in collisions
at smaller radii with larger pion production efficiency. The conversion
to high energy pions 
of a significant fraction, $\ge20\%$, of the energy of protons accelerated
to energy similar to or larger than the break energy $\epsilon_{pb}$,
would be avoided only if $\Gamma\gg100$.

\paragraph*{Neutrino spectrum and flux.}

Roughly half of the energy lost by protons goes into $\pi^0$~'s and the 
other half to $\pi^+$~'s.  Neutrinos are produced by the decay of $\pi^+$'s, 
$\pi^+\rightarrow\mu^++\nu_\mu
\rightarrow e^++\nu_e+\overline\nu_\mu+\nu_\mu$ [the large optical
depth for high energy $\gamma$'s from $\pi^0$ decay, cf. eq. (\ref{tau}),
would not allow these photons to escape the wind]. 
The mean pion energy is $20\%$ 
of the energy of the proton producing the pion. This energy is roughly
evenly distributed between 
the $\pi^+$ decay products. Thus, approximately half 
the energy lost by protons of energy $\epsilon_p$ is converted to neutrinos 
with energy $\sim0.05\epsilon_p$. Eq. (\ref{fpi}) then implies that 
the spectrum of neutrinos above $\epsilon_{\nu b}=0.05\epsilon_{pb}$
follows the proton spectrum, and is harder
(by one power of the energy) at lower energy. 
The break energy is $\epsilon_{\nu b}^{\rm ob.}=5\times10^{14}\Gamma_{300}^2
(\epsilon_{\gamma b,{\rm MeV}}^{\rm ob.})^{-1}{\rm eV}$. For a power law
differential spectrum of accelerated protons $n(\epsilon_p)\propto
\epsilon_p^{-2}$, as typically expected for Fermi acceleration and which
would produce the observed spectrum of ultra-high energy cosmic rays
\cite{CRflux}, the differential neutrino spectrum is $n(\epsilon_\nu)
\propto\epsilon_\nu^{-\alpha}$ with $\alpha=1$ below the break and 
$\alpha=2$ above the break. The spectrum may be modified above the energy
where $\overline\nu_\mu$ are produced by decay of muons with life 
time $\Gamma_\mu t_\mu$ (where $\Gamma_\mu$ is the muon Lorentz factor and 
$t_\mu=2\times10^{-6}{\rm s}$ its rest life time) comparable to
the wind dynamical time $r_d/\Gamma c$, 
i.e. above $\epsilon_{\nu\mu}^{\rm ob.}=
m_\mu c^2r_d/3t_\mu c\approx\Gamma^2m_\mu c^2\Delta t/3t_\mu=
5\times10^{15}\Gamma_{300}^2\Delta t_{\rm ms}{\rm eV}$.
Muons producing $\overline\nu_\mu$ with higher energy may adiabatically
lose a significant part of their energy before decaying (Other energy loss
processes, e.g. inverse-Compton scattering, can be shown to be less important).

Most of the neutrino energy is carried by neutrinos with energy close to the
break energy, $\epsilon_{\nu b}^{\rm ob.}\sim10^{14}{\rm eV}$. The neutrino
flux depends on the relative efficiency with which the wind kinetic energy
is converted to accelerated protons, compared to the efficiency with which 
energy is converted to accelerated electrons (and therefore to $\gamma$-rays). 
If cosmological GRBs are the sources of ultra-high energy cosmic rays,
then the efficiency of converting energy to $\gamma$-rays and 
to accelerated protons should be similar.
The energy production rate required to produce the observed
flux of ultra-high energy cosmic-rays, assuming that the sources are
cosmologically distributed, is $\sim4\times10^{44}{\rm erg\ Mpc}^{-3}
{\rm yr}^{-1}$ over the energy range $10^{19}$--$10^{21}{\rm eV}$,
comparable to the rate of energy production by GRBs as $\gamma$-rays
in the BATSE range \cite{CRflux}. 
For a proton generation spectrum $n(\epsilon_p)\propto
\epsilon_p^{-2}$, similar energy is contained in equal logarithmic
energy intervals, implying that the total energy converted to
accelerated protons (i.e. over the energy range $\Gamma_p m_pc^2
\sim10^{12}$--$10^{21}
{\rm eV}$) is a few times that converted to $\gamma$-ray energy in the BATSE
range. We therefore assume below that the conversion efficiency is
similar for electrons and protons, leading to energy production rate
above the proton energy break, $\sim10^{16}{\rm eV}$, $\dot E\sim
4\times10^{44}{\rm erg\ Mpc}^{-3}{\rm yr}^{-1}$ ($\dot E$ depends
only logarithmically on the value of the break energy). 

The present day neutrino energy density due to GRBs
is approximately given by $U_\nu\approx
0.5f_\pi(\epsilon_{pb})
t_H\dot E$, where $t_H\approx10^{10}{\rm yr}$ is the Hubble time. 
The neutrino flux is therefore approximately given by
\begin{eqnarray}
J_\nu(\epsilon_\nu>\epsilon_{\nu b})\approx&&{c\over4\pi}{U_\nu\over
\epsilon_{\nu b}}
\approx10^{-13}{f_\pi(\epsilon_{pb})\over0.2}
\dot E_{44}\nonumber\\
&&\times\left({\epsilon_{\nu b}^{\rm ob.}\over10^{14}{\rm eV}}\right)^{-1}
{\rm cm}^{-2}{\rm s}^{-1}{\rm sr}^{-1}\,,
\label{Jnu}
\end{eqnarray}
where $\dot E=10^{44}\dot E_{44}{\rm erg\ Mpc}^{-3}{\rm yr}^{-1}$.
The high energy neutrinos predicted in the dissipative wind model of GRBs
may be observed by detecting the Cherenkov light emitted by high energy muons
produced by neutrino interactions below a detector on the surface of the
Earth (see \cite{Gaisser} for a recent review). This technique works only for
muons entering the detector from below, ``upward moving muons'', due to
the large background of downward atmospheric muons.
The probability $P_{\nu\mu}$ that a neutrino
would produce a high energy muon in the detector is approximately given by 
the ratio of the high energy muon range to the neutrino mean free path.
At the high energy we are considering, 
$P_{\nu\mu}\simeq10^{-6}(\epsilon_\nu/1{\rm TeV})$ \cite{Gaisser}.
Using (\ref{Jnu}), the expected flux of upward moving muons is
\begin{equation}
J_{\mu\uparrow}\approx50{f_\pi(\epsilon_{pb})\over0.2}
\left({\dot E\over4\times10^{44}{\rm erg\ Mpc}^{-3}{\rm yr}^{-1}}
\right){\rm km}^{-2}{\rm yr}^{-1}\,.
\label{Jmu}
\end{equation}
The rate is almost independent of $\epsilon_{\nu b}$, due to the increase 
of $P_{\nu\mu}$ with energy. 

The rate (\ref{Jmu}) is comparable to
the background expected due to atmospheric neutrinos \cite{Gaisser}.
However, neutrino bursts should be easily detected above the background, 
since the
neutrinos would be correlated, both in time and angle, with the GRB
$\gamma$-rays. A ${\rm km}^2$ neutrino detector should detect each year
$\sim10$ to 100 neutrinos correlated with GRBs. Furthermore, nearby bright
GRBs, although rare ($\sim0.1$ per year), 
would produce a burst of several neutrinos in a ${\rm km}^2$
detector. A burst at a distance of $100{\rm\ Mpc}$ producing 
$0.4\times10^{51}{\rm erg}$ in $\sim10^{14}{\rm eV}$ neutrinos
would produce $\sim3{\rm\ km}^{-2}$ upward muons.

\paragraph*{Implications.}

Detection of neutrinos from GRBs would corroborate the cosmological fireball 
scenario for GRB production (acceleration of protons to ultra-high 
energy by the processes discussed above is not possible if GRBs are
Galactic \cite{coronal}). Neutrinos from GRBs could be used to
test the simultaneity of
neutrino and photon arrival to an accuracy of $\sim1{\rm\ s}$
($\sim1{\rm\ ms}$ for short bursts), checking the assumption of 
special relativity
that photons and neutrinos have the same limiting speed
[The time delay for neutrino of energy $10^{14}{\rm eV}$ 
with mass $m_\nu$ traveling $100{\rm\ Mpc}$ is only 
$\sim10^{-11}(m_\nu/10{\rm\ eV})^2{\rm s}$].
These observations would also test the weak
equivalence principle, according to which photons and neutrinos should
suffer the same time delay as they pass through a gravitational potential.
With $1{\rm\ s}$ accuracy, a burst at $100{\rm\ Mpc}$ would reveal
a fractional difference in limiting speed 
of $10^{-16}$, and a fractional difference in gravitational time delay 
of order $10^{-6}$ (considering the Galactic potential alone).
Previous applications of these ideas to supernova 1987A 
(see \cite{John} for review), where simultaneity could be checked
only to an accuracy of order several hours, yielded much weaker upper
limits: of order $10^{-8}$ and $10^{-2}$ for fractional differences in the 
limiting speed \cite{c} and time delay \cite{WEP} respectively.

The model discussed above predicts the production of high energy
muon and electron neutrinos
with a 2:1 ratio. If vacuum neutrino oscillations occur in nature, 
then neutrinos that get here should be almost equally distributed between
flavors for which the mixing is strong.
In fact, if the atmospheric neutrino anomaly has the explanation it is
usually given, oscillation to $\nu_\tau$'s with mass $\sim0.1{\rm\ eV}$
\cite{atmo}, then
one should detect equal numbers of $\nu_\mu$'s and $\nu_\tau$'s. 
Upgoing $\tau$'s, rather than $\mu$'s, would be a
distinctive signature of such oscillations. 
Since $\nu_\tau$'s are not expected to be produced in the fireball, looking
for upgoing $\tau$'s would be an ``appearance experiment''
($\nu_\tau$'s may be produced by photo-production of charmed mesons; 
However, the high photon threshold, $\sim50{\rm GeV}$, and
low cross-section, $\sim1\mu{\rm b}$ \cite{charm}, for such reactions imply 
that the ratio of charmed meson to pion production is $\sim10^{-4}$).
To allow flavor change, the difference in squared neutrino masses, 
$\Delta m^2$, should exceed a minimum value
proportional to the ratio of source
distance and neutrino energy \cite{John}. A burst at $100{\rm\ Mpc}$ 
producing $10^{14}{\rm eV}$ neutrinos can test for $\Delta m^2\ge10^{-16}
{\rm eV}^2$, 5 orders of magnitude more sensitive than solar neutrinos.
Note, that due to the finite pion life time, flavor mixing would be caused by
de-coherence, rather than by real oscillations, for neutrinos with masses
$>0.1{\rm eV}$.

The $\nu_\mu$'s may resonantly oscillate to $\nu_e$'s (the MSW
effect) as they escape the fireball, provided the electron number density
exceeds the MSW resonance density (e.g. \cite{John}). 
The fireball electron density is approximately 
$n_e\sim L/4\pi r_d^2\Gamma^2m_pc^3\sim10^{10}$--$10^{12}{\rm cm}^{-3}$, 
implying that the MSW effect could be important only for neutrino
masses smaller than $\sim10^{-12}{\rm eV}^2$. For such low masses, the 
spatial scale of the fireball implies that $n_e$ changes too rapidly
to allow the use of the adiabatic approximation, and that the
influence of the MSW effect would depend on the detailed structure of the
fireball.

\paragraph*{Acknowledgments.} 

We thank J. Appel, J. Butler, T. Gaisser,
P. Krastev, J. Miralda-Escud\'e, B. Paczy\'nski, J. Peoples, J. Rosner,
T. Stanev, D. Seckel, F. Wilczek and M. Witherell
for helpful discussions. This research
was partially supported by a W. M. Keck Foundation grant 
and NSF grant PHY 95-13835.

\end{document}